\documentclass[12pt]{article}
\usepackage{amssymb}
\usepackage{graphicx}
\usepackage[cp1251]{inputenc}
\usepackage{rotating}

\begin{document}
 \noindent {\footnotesize\it Astronomy Reports, 2025, Vol. 69, No. 9, pp. 786--797}
 \newcommand{\dif}{\textrm{d}}

 \noindent
 \begin{tabular}{llllllllllllllllllllllllllllllllllllllllllllll}
 & & & & & & & & & & & & & & & & & & & & & & & & & & & & & & & & & & & & & &\\\hline\hline
 \end{tabular}

  \vskip 0.5cm
  \bigskip
 \bigskip
\centerline{\large\bf Place of the Radcliffe Wave in the Local System}
 \bigskip
 \bigskip
  \centerline { 
   V. V. Bobylev\footnote [1]{bob-v-vzz@rambler.ru},  N. R. Ikhsanov, and A. T. Bajkova}
 \bigskip
 \centerline{\small\it Pulkovo Astronomical Observatory, Russian Academy of Sciences, St. Petersburg, 196140 Russia}
 \bigskip
 \bigskip
{{\bf Abstract}--A review of publications devoted to the study of the characteristics of the Radcliffe wave has been
given. The advent of mass measurements of radial velocities of stars has recently led to a number of interesting
results obtained from the analysis of spatial velocities of stars and open star clusters. An important place in
the study has been given to issues related to the clarification of the direct or indirect influence of magnetic
fields on the process of formation of the Radcliffe wave. The hypothesis of Parker instability of the galactic
magnetic field as one of the reasons for the formation of wave-type inhomogeneities in the galactic disk has
been discussed.
 }

\bigskip
\section{INTRODUCTION}
The Local System is known [1--3] as a complex structural formation near the Sun that fills the space
between two segments of the Carina--Sagittarius and Perseus spiral arms. From a modern point of view, it
includes clouds of neutral (for example, the Lindblad ring [4, 5]) and molecular hydrogen, interstellar dust,
protostars, as well as young stars of various spectral classes. The stars are concentrated mainly in a few
large OB associations [6] and open star clusters, although there are also field stars [7]. The Gould Belt
[7--11], with a radius of $\sim$500 pc, is part of the Local System. A number of authors [10, 11] identify the
Local System with the Orion Arm. Thus, the size of the Local System is $\sim$1 kpc in the diameter (along the
galactic axis ), and it is extended by 3--4 kpc in the perpendicular direction (approximately along the axis $y$).

The publication of a catalogue of molecular clouds [12] with highly accurate estimates of their distances
(with an average error of $\sim$5\%) became an important milestone in the study of the Local System. In particular,
based on the analysis of this catalogue, the Radcliffe wave was discovered [13], which is a narrow
chain of clouds, stretched almost in a line with a length $\sim$2.7~kpc in the galactic plane $xy$. The main feature of
this structure is the clearly visible wave-like nature of the distribution of clouds in the vertical direction. In
this case, the maximal value of the coordinate of $z\sim160$~pc is observed in close proximity to the Sun. According to its discoverers [13], the wave is attenuating. These authors also concluded that the narrow chain distribution of molecular clouds refutes Blaauw's [14] hypothesis that the Gould Belt could have formed as a result of a hypernova explosion.

In the hypernova explosion model, the shock wave in the galactic disk forms an expanding ring-shaped
structure, which, due to the galactic tide, is stretched over time along its galactic orbit and takes on an elliptical
shape. This model explained well the existence of the Lindblad ring. However, this model is no longer able to explain the significant tilt of the Gould Belt to the galactic plane.

There is currently no clear understanding of the nature of the Radcliffe wave. According to [15], the
origin of the Radcliffe wave is associated with the Kelvin--Helmholtz instability, which occurs in the disk of
the Galaxy because of the difference in the rotation speeds of the dark matter halo and the disk. A hypothesis
is being discussed about the impact of an external impactor on the galactic disk [16], which could be a
dwarf galaxy---a satellite of the Milky Way, a massive clot of dark matter, or a globular cluster.

According to a number of authors [17, 18], the Radcliffe wave could have arisen as a result of the
impact of shock waves from several supernovae and their stellar winds during the formation of the Local
Bubble or North Polar Spur. However, the most important question about the role that the magnetic
field plays in the formation of bubbles remains open to this day [19].

Currently, more than a dozen publications have been devoted to the study of the Radcliffe wave. The
wave-like behavior of vertical coordinates has been confirmed in the distribution of interstellar dust [20, 21], molecular clouds [22], masers and radio stars [23], T Tauri stars [24], massive OB stars [16, 25], and young open star clusters [18, 25, 26].

This paper is aimed to review the most important aspects related to the study of the Radcliffe wave.
Since the publication of our review on this topic [27], a number of significantly new results have emerged
[17, 18, 21, 28--33]. The nature of the distribution of object coordinates in the Radcliffe wave is relatively
well known. The properties of their vertical velocities have been studied significantly less, and data for analysis
have only become available in very recent times. An important place in this paper is given to issues not previously addressed, related to the clarification of the possible direct or indirect influence of magnetic fields on the process of Radcliffe wave formation.

 \section{PROPERTIES OF THE RADCLIFFE WAVE}
In this paper, we consider the heliocentric rectangular coordinate system $xyz,$ in which the $x$ axis directed from the Sun to the center of the Galaxy, the direction of the $y$ axis coincides with the direction of rotation of the Galaxy, while the $z$ axis is directed
towards the north galactic pole, as well as the galactocentric rectangular coordinate system $XYZ$, in which
the $X$ axis directed from the center of the Galaxy to the Sun, the direction of the $Y$ axis coincides with the
direction of rotation of the Galaxy, and the $Z$ axis directed towards the north galactic pole. Thus, in these two coordinate systems, only the directions of the $X$ axes and differ. In this case, the orientation of the Radcliffe wave with respect to the and axes differs
in the sign only. For example, in the heliocentric coordinate system, the transition to the hatched axis by the angle of 30$^\circ$ is carried out as follows:
\begin{equation}
 y'= y\cos{30^\circ}+x\sin{30^\circ}.
 \label{y'-30}
\end{equation}

\begin{figure}[t]{ \begin{center}
  \includegraphics[width=0.95\textwidth]{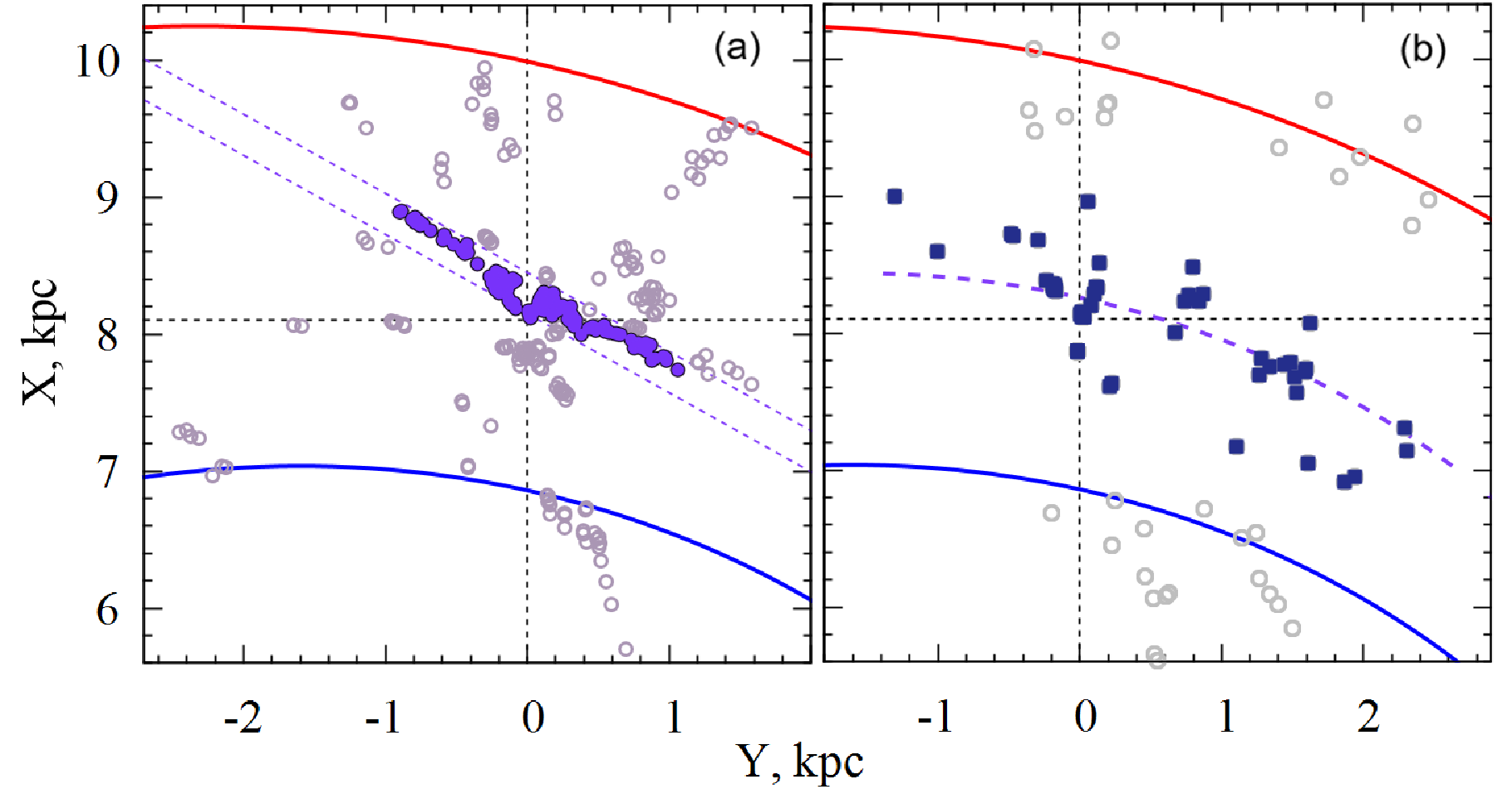}
  \caption{
  (a) Distribution of 380 molecular clouds from [13] projected onto the galactic plane $XY$ (gray circles) and 189 clouds from
a narrow zone (limited by blue dotted lines) passing at an angle of $-30^\circ$  to the $Y$axis (blue filled circles); (b) distribution of maser sources with measured trigonometric parallaxes, Orion arm masers are shown as dark squares. Two fragments of a four-arm spiral pattern with a twist angle $i=-13^\circ$ of are noted, the red line is a segment of the Perseus arm, the blue line is a segment of the Carina--Sagittarius arm, and the Sun is located at the point with coordinates of $(X,Y)=(8.1,0)$~kpc.
 }
 \label{Alves-XY}\end{center}}\end{figure}
\begin{figure}[t]{ \begin{center}
  \includegraphics[width=0.95\textwidth]{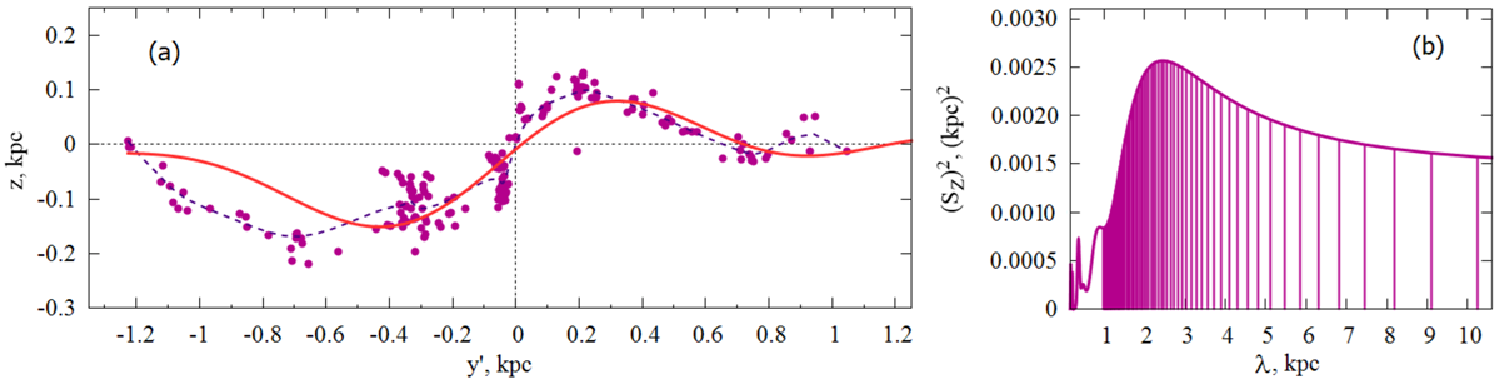}
  \caption{(a) Vertical coordinates of a sample of molecular clouds as a function of distance $y'$; the periodic bold (red) line reflects
the results of spectral analysis, while the dashed (blue) line shows the smoothed average values of the coordinates. (b) Power spectrum of a sample of molecular clouds. The figure is taken from the study of Bobylev et al. [27].
  }
 \label{Alves-poly-RW}\end{center}}\end{figure}
\begin{figure}[t]{ \begin{center}
  \includegraphics[width=0.75\textwidth]{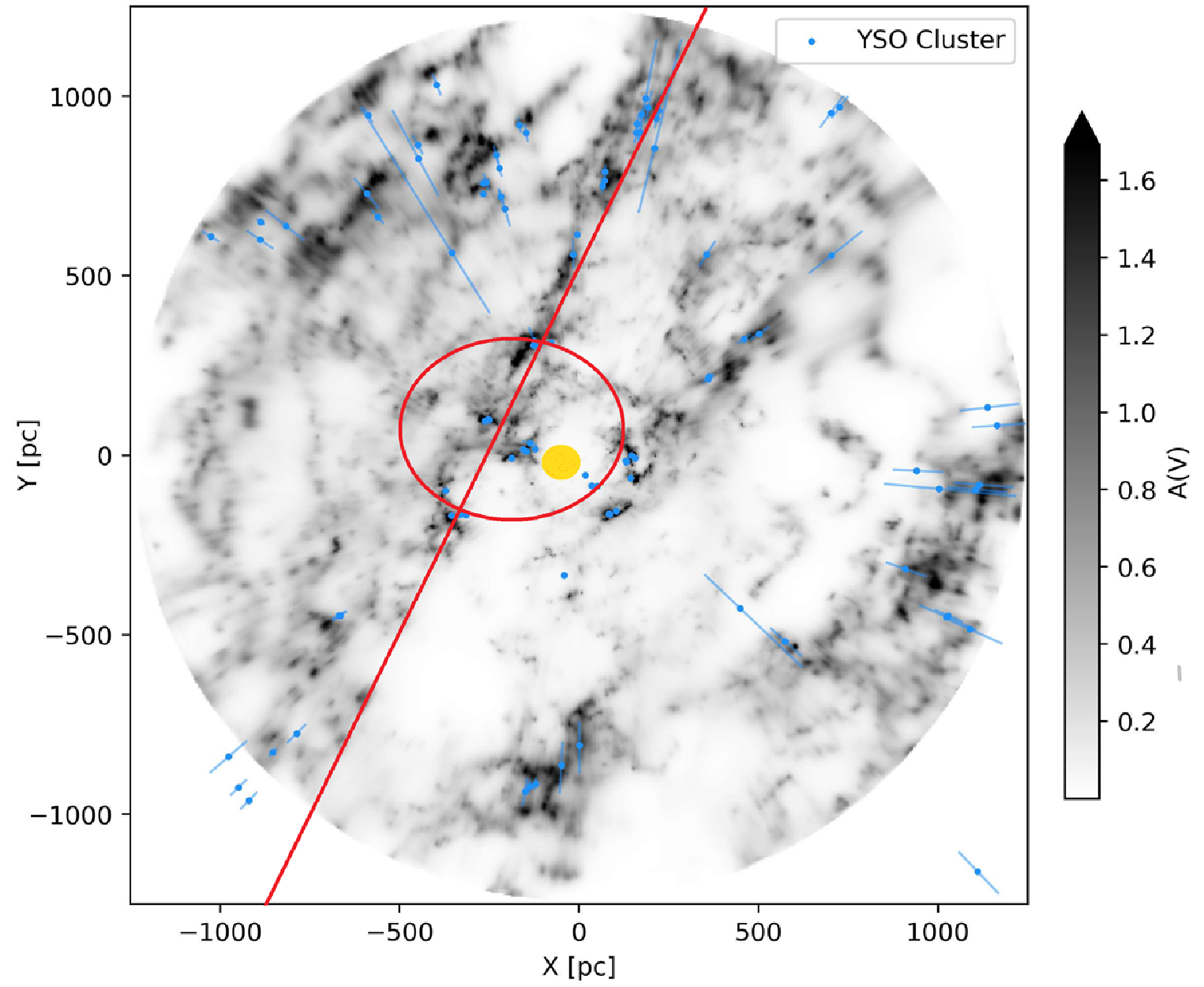}
  \caption{Distribution map of dust matter from [21] with the condensations of young stellar objects (YSOs) plotted, on which we
plotted the position of the Sun (yellow circle), a line with a slope of $25^\circ$ to the $Y$ axis (red line) and the approximate position of the Gould Belt (red circle).
 }
 \label{f-Edenh}\end{center}}\end{figure}
\begin{figure}[t]{ \begin{center}
  \includegraphics[width=0.75\textwidth]{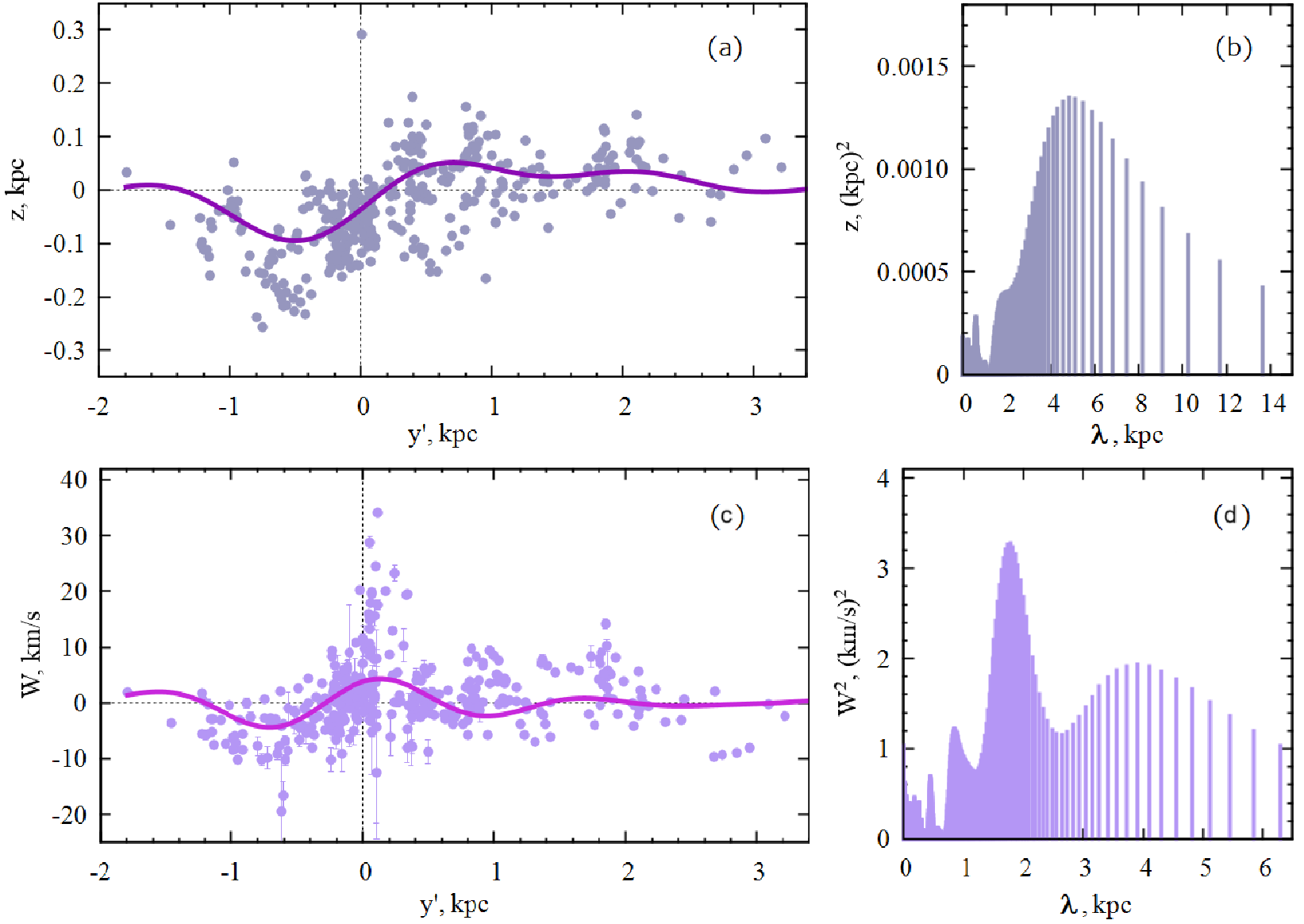}
  \caption{(a) Vertical coordinates of young OSCs depending on the coordinate $y'$, (b) their power spectrum, (c) vertical velocities
of OSC depending on the coordinate $y'$, and (d) their power spectrum. The figure is taken from [43].
  }
 \label{f-RAA}\end{center}}\end{figure}
\begin{figure}[t]{ \begin{center}
  \includegraphics[width=0.8\textwidth]{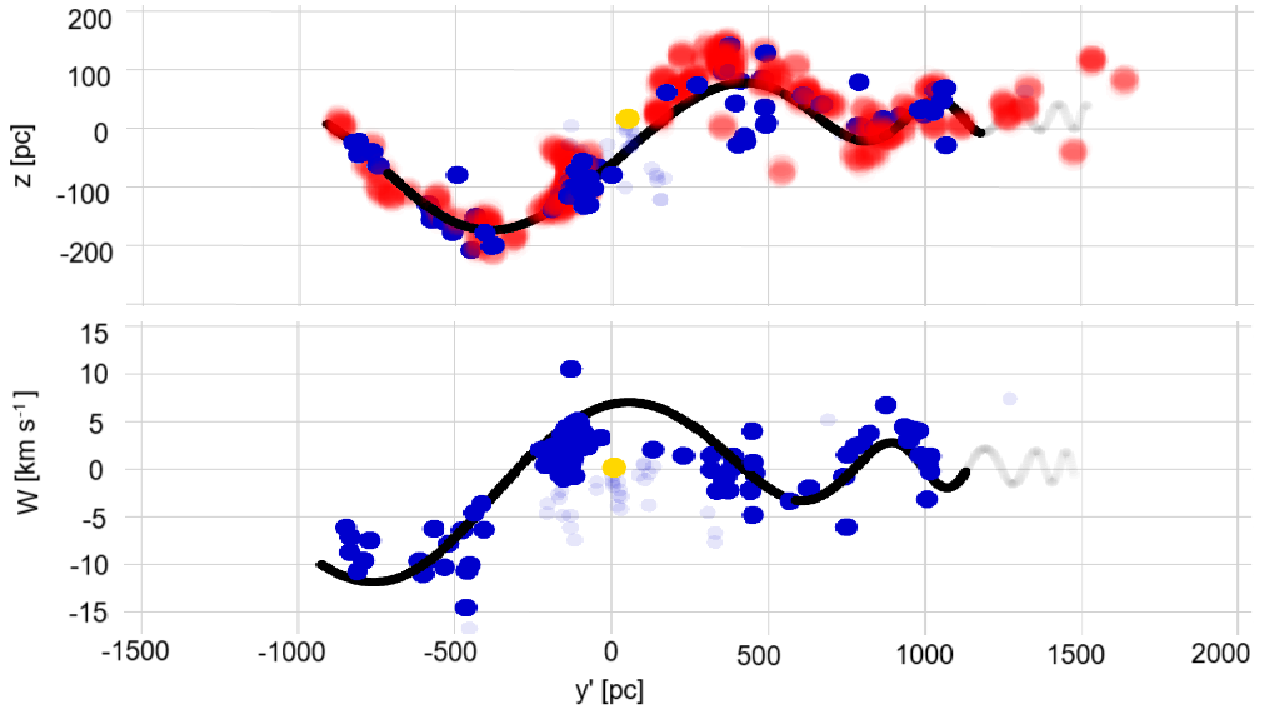}
  \caption{
  Vertical coordinates of young OSCs (upper panel) and their vertical velocities (lower panel) as a function of coordinate $y'$
that were obtained in [18].
 }
 \label{f-Kon-Z-Vz}\end{center}}\end{figure}
\begin{figure}[t]{ \begin{center}
  \includegraphics[width=0.85\textwidth]{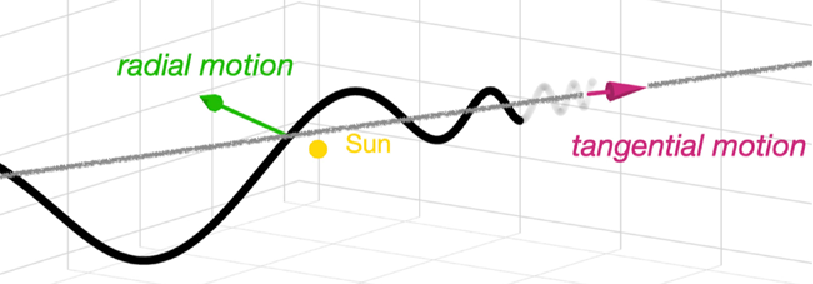}
  \caption{
Radial (towards the galactic anticenter) and tangential motion of the Radcliffe wave according to [18], and the $y'$ axis
shown by the gray line.
 }
 \label{f-Kon}\end{center}}\end{figure}

 \subsection{Molecular Clouds}
The detection of the Radcliffe wave was made possible by the studies [12, 34], where distances to 326 molecular clouds concentrated in approximately 60 regions of active star formation were obtained were estimates. The clouds are located in the Orion Arm region at heliocentric distances from 150 pc to 2.5 kpc with altitudes of $|z|<300$~pc. These authors developed a method [34] that combines high-precision photometric data with trigonometric parallaxes of stars from the Gaia DR2 catalogue [35]. According to
their estimates, the distances to molecular clouds were determined with an average error of $\sim$5\%.

According to the data from the catalogue [12], in [13], a narrow chain of molecular clouds that is stretched out almost in one line that is located at an angle of $\sim$30$^\circ$ to the galactic $y$ axis , was identified. In vertical coordinates $z$, a wave of clouds was discovered. The first authors of this scientific team are representatives of the Radcliffe Institute for Advanced Study in Cambridge, Massachusetts, so they named the wave in honor of their home institute.

Actually, it is not entirely correct to call this formation a wave. In fact, the Radcliffe wave is a periodicity
with variable amplitude. Its main feature is the indication of the presence of a strong disturbance of the vertical
coordinates of molecular clouds.

The Radcliffe wave modeling in [13] was performed using a quadratic function in coordinate space $xyz,$
that is defined by three sets of ``reference points'' $(x_0,y_0,z_0),$ $(x_1,y_1,z_1),$ and $(x_2,y_2,z_2).$  The wave-like
behavior about the wave center was described by a sinusoidal function about the plane $XY$ with a decaying
period and amplitude. As a result, Alves et al. [13] found the following parameters of the model wave:
 \begin{equation}
 \renewcommand{\arraystretch}{1.0}
 \label{rez-Alves}
 \begin{array}{lll}
  \lambda=2.7\pm0.2~\hbox{kpc},\\
  A=160\pm30~\hbox{pc},\\
  \sigma_{\rm scatter}= 60\pm15~\hbox{pc},\\
  \hbox{Mass}\geq 3\times 10^6 M_\odot,
 \end{array}
 \end{equation}
where $\lambda$ is the wavelength, $A$ is the wave amplitude, $\sigma_{\rm scatter}$ is the standard deviation of the vertical coordinates of clouds from the model, and the given errors of the parameters in (2) correspond to the 95\% level ($\pm 2\sigma$).

In the study of Bobylev et al. [27], a Fourier analysis was applied to the same molecular cloud data, from which the Radcliffe wave was discovered in [13]. For this purpose, 189 clouds located in a narrow zone inclined at an angle of 30$^\circ$ to the $y$ axis were selected. The distribution of these clouds in projection onto the galactic plane $XY$ is given in Fig. 1.

For the case of a polychromatic wave (when all frequencies of the main lobe in the power spectrum are
used), it was found:
 \begin{equation}
 \label{rez-Z-poly}
 \begin{array}{lll}
  \lambda=2.5\pm0.1~\hbox{kpc},\\
  z_{max}=150\pm4~\hbox{pc},\\
  \sigma_z=46~\hbox{pc},
 \end{array}
 \end{equation}
where the $\lambda$ value corresponds to the maximum of the power spectrum,  $z_{max}$ corresponds $A$ in (2),  
$\sigma_z$ corresponds $\sigma_{\rm scatter}$ in (2), and the parameter errors correspond to the level of 68\% ($\pm1\sigma$). The result (3) is reflected in Fig. 2, from which it is evident that the maximal value of the wave $z_{max}$ is achieved at $-0.4$~$y'\approx-0.4$~kpc. The main lobe frequencies of the power spectrum used to construct the periodic curve are
marked by vertical lines in Fig. 2b. Thus, we have good agreement between the values of parameters (2) and
(3) that are obtained by different methods.

We note the study [28], which presents a new catalog of distances to molecular clouds located within $\sim$2.5~kpc from the Sun. Distances to 63 clouds were obtained using parallaxes of $\sim$3000 young stars from the Gaia DR3 catalogue [36]. According to [28], the typical uncertainty in measuring the distance to clouds is approximately 3\% (internal error). Good agreement
was found with a spread of $\sim$10\% between the new distances to the clouds and previous estimates (for example,
with the estimates from [12]). According to [28], unlike the distances to clouds obtained by taking into account interstellar absorption, this catalog is a more suitable reference material for studying the physical properties of nearby dense regions. Although a significantly smaller number of clouds was used compared to [13], the presence of disturbances in the vertical positions
of clouds that is characteristic of the Radcliffe wave was confirmed.

 \subsection{Interstellar Dust}
The nature of the distribution of gas and dust clouds in the Local System is of great importance for
the study of a number of astrophysical problems, in particular, for the analysis of the spiral structure, as
well as the study of the properties of the Radcliffe wave.

In [20], photometric data on stars from the Gaia EDR3 catalogue [37] were combined with infrared
photometric data on stars from the 2MASS catalogue [38]. Based on this material, a three-dimensional map
of interstellar absorption in the local volume was constructed. The vertical sections show wave-like deviations
from the galactic plane with an amplitude of up to 300 pc in different directions. While in section number
3 that is oriented in the direction of $l=60^\circ$, the manifestation of the Radcliffe wave is clearly visible.

Edenhofer et al. [21] presented a new high-resolution 3D dust distribution map that extends to 1.25 kpc
from the Sun. According to these authors, the reconstruction has a high dynamic range and reveals weak
dust lanes in the considered region. Small, roughly
spherical cavities are visible throughout the map. The dust clouds are compact and only weakly elongated
radially. According to these authors, the Radcliffe wave is resolved at an unprecedented level of detail
that is previously only available for the closest dust clouds.

The map from [21] is available online\footnote{https://www.aanda.org/articles/aa/olm/2024/05/aa47628-23/aa47628-23.html}. It is presented in Fig. 3, where the condensations of young stellar objects (YSOs) are plotted. For this purpose,
data from catalogues [39--41] were used. As can be seen from the figure, the YSOs that we associate with
the Radcliffe wave are located along the red line. At this angle, we set the selection zone in [23] when analyzing
the YSO sample from the catalog [42].

 \subsection{Young Stars}
Young stars are of great interest for studying the properties of the Radcliffe wave, since for many of
them, besides their exact positions, their spatial velocities are known. Here, we will consider them in order
of increasing age: stars that have not reached the main sequence stage (YSOs, young stellar objects); masers
and radio stars, among which are both protostars and stars that have not reached the main sequence stage, as
well as several massive stars no older than OB2; and finally, OB stars.

 \subsubsection{YSO}
Li and Chen [24] found a relationship between their vertical positions and vertical velocities
using data from YSOs tracing the Radcliffe wave. The
vertical velocities of the stars were calculated without using radial velocities, only from their proper motions.
Despite this, these authors obtained an important result. Namely, they found a phase difference of
$\sim2\pi/3$ between waves in coordinates and velocities.

In the study of Bobylev et al. [23], a sample of YSOs from the catalogue [42] was studied. Based on a
Fourier analysis of 600 stars that have not reached the main sequence stage, an estimate of the amplitude
$z_{max}=118\pm3$~pc (which is achieved when $y'=-0.4$~kpc) and wavelengths of $\lambda=2.0\pm0.1$~kpc was
obtained, which is in good agreement with the values of parameters (2) and (3). Unfortunately, for 90\% of
the selected stars, there were no radial velocity data, only their trigonometric parallaxes and proper
motions from the Gaia DR2 catalogue were available [35].

 \subsubsection{Masers and radio stars}
In the study of Bobylev et al. [23], in addition to YSO, a sample of maser sources and radio stars with trigonometric parallaxes measured by the VLBI method was also studied. 68 such objects were used with parallax errors less than 15\%. Based on a Fourier analysis, an estimate of the amplitude value was obtained $z_{max}=87\pm4$~pc (which is achieved when $y'=-0.28$~kpc) and wavelengths $\lambda=2.8\pm0.1$~kpc, which is in good agreement with the values of parameters (2) and (3). In this case,
for the first time, reliable estimates of velocity disturbances $W$ in the Radcliffe wave were obtained. The following values of the maximal disturbance $W_{max}$ (which is achieved when $y'=1.4$~kpc) and the wavelengths of these disturbances velocity were found:
 \begin{equation}
 \label{sol-68-masers-W}
 \begin{array}{lll}
   W_{max}= 5.1\pm0.7~\hbox{km/s},\\
   \lambda= 3.9\pm1.6~\hbox{kpc}.
 \end{array}
 \end{equation}

 \subsubsection{OB stars}
In the aspect of studying the properties of the Radcliffe wave, a sample of very young OB stars was first analyzed in [25]. Besides, these authors analyzed 13 open star clusters (OSCs) younger than 30 million years from the solar neighborhood with a radius of $\sim$2 kpc and found that, single OB stars compared to OSCs are worse at representing an elongated structure that traces the Radcliffe wave, so the main emphasis was placed on the analysis of OSCs. Although they showed that there is some relation
between the perturbations of the vertical positions and vertical velocities of the analyzed objects, their results are preliminary due to the low power of the samples.

 \subsection{Open Star Clusters}
For most open star clusters located within the Local System, their heliocentric distances, proper motions, and radial velocities, the values of which are calculated as averages over many members of the cluster, are known. Thus, the value of OSCs is that they
allow for a highly accurate analysis of both their positions and their spatial velocities. Precisely, for the OSCs, new, interesting results describing the properties of the Radcliffe wave have now been obtained.

In [43], a sample of 374 OSCs younger than 25 million years (with an average age of 9.7 million years) was analyzed. The data for these clusters were taken from the catalogue [44], one of the most complete to date (it includes more than 7000 OSCs, and the average distances and velocities for them were calculated using data from the Gaia DR3 catalogue [36]). Based on the Fourier analysis, the following estimates were obtained: (a) the maximal value of the vertical coordinate
was $z_{max}=92\pm10$~pc and the wavelength was $\lambda=4.82\pm0.09$~kpc, and (b) the maximal disturbance
velocity was $W_{max}=4.36\pm0.12$~km/s (achieved at $y'=-0.71$~kpc), and the wavelength of these disturbances
was $\lambda=1.78\pm0.02$~kpc. These results are shown in Fig. 4.

The authors of [18] analyzed the OSC parameters taken from the catalog [45] (containing $\sim$1900 OSCs
with astrometric data from the Gaia DR2 catalog [35]) and found that the Radcliffe wave oscillates in the
plane of the Galaxy and also drifts radially from the galactic center. The dependences of vertical coordinates
and vertical velocities on the coordinate $y'$ that are obtained by them from a sample of young OSCs
are shown in Fig. 5. Figure 5 was constructed by us based on the interactive graphic data of these authors\footnote{https://faun.rc.fas.harvard.edu/czucker/rkonietzka/radwave/interactive-figure1.html}.
It is evident that there is a phase difference between the waves in coordinates and velocities, noted earlier in [24].

The diagram of the radial and tangential motion of the Radcliffe wave found in [18] is given in Fig. 6. Besides, these authors concluded that the radial outward drift of the Radcliffe wave from the Galactic center suggests that the cluster whose supernovae eventually created today's expanding Local Bubble may have been born in the Radcliffe wave. Treating the Radcliffe wave as a single coherent structure in space and velocity that responds to the galactic potential, they found that the structure is well modeled as a damped sine wave with a maximal amplitude of $\sim$220 pc and a mean wavelength of $\sim$2 kpc. The corresponding maximal
vertical velocity is $\sim$14 km/s.

By modeling the motion of individual clouds in the
galactic potential, the authors of [32] showed that the
shear and epicyclic motion stretch the Radcliffe wave
almost twice its current length over time on the scale
of $\sim$45 million years. They also found that during this
time, new thread-like structures can form and existing
ones can merge. The figures in their paper clearly show
that a thin chain of clouds in the present time (as the
red line in Fig. 3) very quickly takes on a sawtooth
appearance over time. These authors insist that it is an
oversimplification to regard the Radcliffe wave as a
rigid body in the galactic disk as an oscillating structure
in the vertical direction.

In [31], high-resolution spectroscopy was performed on 53 stars, members of seven OSCs (the age
is $\leq$100 million years) that are located along the Radcliffe wave. Radial velocities and atmospheric parameters
were presented for all 53 stars. For a subsample of FGK stars, abundances for 25 chemical elements were
determined. On average, OSCs show chemical compositions
consistent with those of the Sun. Finally, a
correlation between the chemical composition, age,
and position of the clusters along the Radcliffe wave
that demonstrates the physical connection of OSCs of
different ages within the inhomogeneous mixing scenario
was obtained.

\section{RADCLIFFE WAVE AND GOULD BELT}
The geometric characteristics of the Gould belt, which were determined from the OSCs and from young stars, are well known [14, 46, 47]. As can be seen from Fig. 3b, the center of the Gould belt is located very close to the symmetry axis of the Radcliffe
wave. According to, for example, Fig. 5, it is located in the region where the amplitudes of disturbance
of vertical positions and vertical velocities reach the highest values. All this indicate the common origin of the Radcliffe wave and the Gould belt, and their origin and further evolution are inseparable.

 \begin{table}[t] \caption[]{\small Estimates of the twist angle and width of the Orion arm }
  \begin{center}  \label{t:orion}
  \small
  \begin{tabular}{|c|c|c|c|c|c|}\hline
 \quad $i,$~deg & Width, kpc & Objects & $n$ & Year of publication & Reference \\\hline
 $-13$              &             &            H II zones &   90 &   1970 & [56] \\
 $-10.1\pm2.7$ &             &               masers &   30 &   2013 & [57] \\
 $-12.9\pm2.9$ &             &               masers &   23 &   2014 & [58] \\
 $-11.4\pm1.9$ &   0.31    &             masers &   28 &   2019 & [51] \\
 $~-9.9\pm1.2$ &            & young Cepheids & 140 &   2020 & [59] \\
 $-11.3\pm1.9$ &   0.30    &             masers &   28 &   2023 & [60] \\
 $-10.0\pm2.5$ &   0.26    &          OB2 stars & 2363&   2023 & [60] \\
 \hline
\end{tabular}\end{center} \end{table}

\section{RADCLIFFE WAVE AND AND ORION ARM}
\subsection{Orion Arm and Grand Design Structure}
Already in Fig. 1a, the absence of any chain of clouds located at an angle of 10$^\circ$--15$^\circ$ to the axis ,
characteristic of the Orion arm is striking.

Twist angle $i$ of the Orion arm has been repeatedly estimated by various authors using various data. A
brief summary of these determinations is given in Table 1. It should be noted that, according to modern
estimates, the Orion arm is not part of the global galactic spiral structure grand design, but is a branch
(spur) from some arm of this structure. However, at present, it is not yet clear from which one specifically
is: from the Carina--Sagittarius arm or Perseus. On the other hand, the Orion Arm could have formed under
the influence of a spiral density wave. Therefore, following the approach of Reid et al. [48], in Fig. 1b, a dashed line is drawn along the masers of the Orion arm and corresponds to a segment of a logarithmic spiral with a twist angle of $-13^\circ$.

It is important to have estimates of the Orion arm twist angle that are not model dependent. It is known
that fitting to observational data (positions and velocyities) based on some model of the spiral pattern that
depends on the number of spiral arms $m$ produces different values of the twist angle.

For example, in [54], when determining the parameters of the galactic spiral pattern, approximately equal $\chi^2$ values were obtained for several solutions both with $m=4$ and with $m=2$. However, for solutions with $m=4$, the twist angle was found to be
equal to $i\sim-26^\circ$ , while for solutions with $m=2$, the twist angle was found to be within the range from $-10^\circ$ 
to $-15^\circ$. Based on the two-arm logarithmic model, in [55], for giant molecular clouds, by H II zones, 2MASS sources, and H I neutral hydrogen clouds, an estimate of $i=-5.56\pm0.06^\circ$ that is the same for both sleeves was obtained. In [56], from the analysis of various indicators of the spiral structure based on the four-arm logarithmic model with $m=4$, a value of $i=-12.8^\circ$
as the most probable was found. In this paper, for the grand design structure, we use a four arm spiral pattern with a twist angle of $i=-13^\circ$ (see Fig. 1) according to [57].

In Table 1, the results obtained by direct methods independent of the number of model spiral arms are reflected. Such methods include, for example, the analysis of a diagram of ${\log (R/R_0)-\theta}$, where $R$ and $R_0$ are the galactocentric distances to the star and the Sun, respectively, while is the position angle of the star. In this case, obviously, the result of the estimate
depends on the number of used sources. For example, in [58], a preliminary estimate of the twist angle of the Orion arm of $-27.8\pm4.7^\circ$, which differs significantly from subsequent estimates by this group of authors obtained using this method on masers, was obtained based on only four masers with measured trigonometric parallaxes.

The papers listed in Table 1 analyzed high-precision positions of very young objects: masers with measured
trigonometric parallaxes using the VLBI method, H II zones, OB2 stars, and young Cepheids. As csn be seen, all the estimates of the twist angle value given in the table are in the range from $-10^\circ$ to $-13^\circ$. These values are very different from the tilt of the chain of molecular clouds of $-30^\circ$, by which the Radcliffe wave was discovered.

Nevertheless, we note a recent paper [59], where the spiral structure of the Galaxy was studied based on a sample of approximately 500 H II zones, the distances to which were determined using trigonometric parallaxes of OB2 stars from the Gaia DR3 catalogue.
These authors concluded that the segments of spiral arms outlined by young objects and evolving stars are
consistent with each other, at least in the solar neighborhood.
In particular, the Local Arm, outlined by young objects is interpreted by them as a segment of an arm with a large twist angle of $-25.2^\circ\pm2.0^\circ$, the inner edge of which is in good agreement with the Radcliffe wave.

\subsection{Large-Scale Waves in the Galactic Disk}
The wealth of information provided by the new stellar data catalogues of the Gaia project has led to the
discovery of various types of flexural waves in the Galactic disk (e.g., [60, 61]).

In this regard, it is interesting to note the study [62], in which the vertical structure and kinematics of
the galactic disk were analyzed using two samples: young giants and classical Cepheids. Based on these
data, a three-dimensional shape of the large-scale disk warp was constructed. They obtain the warp amplitude
of $\sim$0.45~kpc at $R\sim$12~kpc and $\sim$0.7~kpc at $R\sim$14~kpc, which is in full agreement with the estimates
of other authors. These authors then discovered a vertical wave above the warp that propagates radially
from the galactic center. However, they concluded that for a number of reasons, the found wave was different
from the Radcliffe wave. The reasons are as follows: (1) the scale of the Radcliffe wave is significantly
smaller than that found, (2) the waves have significant differences in wavelength, and (3) their orientation
(and position in the galactic disk) are completely different.

\section{OTHER LOCAL SYSTEM DETAILS}
As is already clear from Fig. 3, the Local System (Orion Arm) is not limited to the Gould Belt and the
Radcliffe Wave. In particular, this figure clearly shows the fiber stretched out in length of $\sim$700 pc and located
at an angle of $\sim45^\circ$ to the axis to the right of the Sun, which almost touches the red circle (the boundary of
the Gould Belt). This gas and dust structure is designated as ``Split'' in [63].

In [64], the kinematics of the G120 complex, which they identified using 870-$\mu m$ maps of the Plank
satellite [65], was studied. In Fig. 3, this complex, with
a length of $\sim$500 pc must be located to the left of the
Sun, at the top of the Gould Belt boundary, almost
parallel to the $X$ axis. Lee et al. [64] calculated the
spatial velocities of these clouds by combining the
radial velocities of the gas obtained from $^{12}$CO maps with proper motions of stars associated with gas. Using this data, they modeled the movement of gas clouds in the galactic disk.

In [32, 64], the joint kinematic evolution of three structures: the G120 complex, the Radcliffe wave, and
the ``Split'' was traced based on the assumption that
these three structures form an evolutionary sequence.
The G120 complex is thought to be an early stage of
filament formation, the Radcliffe wave is a typical
example of a galactic-scale filament, while the Split is
a galactic-scale filament at a later stage of evolution.
This conclusion is based on the results of the following
evolutionary prediction: as the simulation progresses,
the G120 complex turns into a thread, while the Radcliffe wave and the    ``Split'' are greatly stretched. Thus, as the simulation showed, the modern G120 gas complex will turn into a filament similar to the Radcliffe wave, while the Radcliffe wave will be stretched.

Mapping the Milky Way's spiral arms vertically is a hallenging problem. The authors of [66] analyzed a  
sample of young giant stars from the Gaia DR3 catalog in order to create a three-dimensional map of
metallicity excess. The constructed map shows traces of spiral arms. The vertical heights of the arms vary
across the galactic disk, reaching amplitudes of up to
400 pc, and exhibit vertical asymmetry relative to the
midplane. In particular, the Perseus arm exhibits high
vertical asymmetry consistent with the galactic warp.
Moreover, evidence of a metal-rich stellar structure
that oscillates vertically almost in phase with the Radcliffe wave has been found. This new structure is larger
in size and extends beyond the Radcliffe wave, reaching vertical amplitudes of $\sim$270 pc and extending at least four kpc in length. For at least half its length, this Radcliffe Extended Wave is the inner edge of the Local Arm.

The solar system revolves around the Milky Way. In doing so, it encounters various galactic environments,
including dense regions of the interstellar medium. These collisions can compress the heliosphere,
exposing parts of the Solar System to the interstellar medium and increasing the influx of interstellar
dust into the Solar System and Earth's atmosphere. The discovery of new galactic structures such as the
Radcliffe wave raises the question of whether the Sun has collided with any of them. In [67], the potential
passage of the Solar System through the Radcliffe wave over the past 30 million years was studied. For
this purpose, a sample of 56 open clusters younger than 30 million years associated with the Radcliffe
wave was used. The trajectory of the Solar System has been shown to have crossed the Radcliffe Wave in the
Orion region 18.2--11.5 million years ago. Moreover, the closest approach occurred 14.8--12.4 million years
ago. This period coincides, on average, with the climate
transition on Earth, which makes it possible to
consider a possible connection with paleoclimatology
and the potential impact of the crossing of the Radcliffe wave on the Earth's climate. As these authors note, such an intersection could lead to anomalies in the content of radionuclides, which is an important research topic in the fields of geology and nuclear
astrophysics.

\section{SEARCHING FOR ANALOGUES OF THE RADCLIFFE WAVE}
May et al. [68] particularly noted a narrow chain of masers with a length of 3--4 kpc extends in the direction $l\sim40^\circ$, passing from a segment of the Carina--Sagittarius spiral arm to the Scutum arm. These authors suggested that this structure may be analogous to the Radcliffe wave.

In Bobylev's study [69], this chain of masers was tested for the presence of periodic perturbations of vertical coordinates and vertical velocities. For this purpose, 65 masers with estimated trigonometric parallaxes
of less than 15\% were selected in a narrow zone with a width of $\sim$1 kpc. As a result, no significant periodic
perturbations of vertical coordinates and velocities were found in this structure characteristic of the
Radcliffe wave. However, in the course of the analysis,
it was discovered that the chain of masers, which
already extends from the Sun (capturing the Gould Belt region) to the section of the Scutum arm, looks
like a section of a logarithmic spiral with a twist angle of $-48^\circ$. It has been suggested that such a large-scale
density wave could have triggered the formation of the Gould Belt, as well as excited the wave in vertical coordinates
and velocities in the Radcliffe wave. It would be interesting to test this hypothesis on more material.
At present, high-precision parallax measurements are lacking to verify it.

\section{MAGNETIC FIELDS}
The results of the reconstruction of the interstellar
magnetic field obtained based on polarimetric observations
of stars located in the vicinity of the Sun were
recently presented in [30]. Their analysis revealed a
clear correspondence between the direction of the
interstellar magnetic field lines and the direction of
the projection of the Radcliffe wave onto the plane of
the sky for stars located within 400 pc of the Sun.
However, the opposite situation is found for stars located at distances exceeding 1.2 kpc. The force lines of the interstellar magnetic field at these distances are located predominantly in the plane of the galactic disk.

In the vicinity of the Sun with a radius of $\sim$1 kpc, the lines of force of the interstellar magnetic field
exhibit a noticeable inclination to the plane of the galactic disk. The average value of the slope is of $\sim18^\circ$.
In the direction where the Radcliffe wave crosses the central plane of the galactic disk, the magnitude of the
tilt angle, however, increases to $30^\circ$. The same angle of inclination to the plane of the galactic disk in this
region is also found in the disturbed part of the gas and dust structure that is stretched along the magnetic
field lines and constitutes the Radcliffe wave. Gas movement is also observed, apparently in the direction
of the magnetic field lines. The velocity of gas movement along the Radcliffe wave increases as the wave approaches the plane of the galactic disk and in the region of their intersection reaches its maximum, which is $\sim$10 km/s.

The totality of the obtained data allowed the authors [30] to consider a hypothesis, in which the
formation of the Radcliffe wave occurred as a result of the instability of the galactic magnetic field. One possible
scenario is the so-called Parker instability (see [70] and the references therein). The change in the configuration of the interstellar magnetic field in this case occurs as a result of the emergence of magnetic field tubes in a direction perpendicular to the plane of
the galactic disk. In the context of this scenario, magnetic
field disturbances lead to a spatial redistribution of the predominantly gas--dust component, which is
in a state close to being frozen into the interstellar magnetic field. The emergence of the magnetic field
leads to the formation of an arched structure, the scale of which in the direction perpendicular to the galactic
plane is of the order of the thickness of the galactic
disk. The characteristic size of the arch in the direction
along the magnetic field lines in this case corresponds
to the wavelength of the Parker instability, which in the
considered case turns out to be of the order of 1--2 kpc [71--73], which corresponds to the observed Radcliffe
wavelength. The characteristic time of development of the Parker instability and formation of a wave-like
structure for the physical conditions realized in the
Local Group is on the order of several tens of millions
of years. Estimation of this parameter is complicated
by the pressure created by cosmic rays, which are one
of the most significant destabilizing factors. The latest
research results obtained in this direction show that
the Parker instability can develop on a time scale of $\sim$20 million years [74]. The relaxation of such a structure is caused by the process of magnetic field diffusion, the characteristic time of which significantly (by a couple of orders of magnitude) exceeds the time of instability development.

It should be noted that Parker instability is a mechanism capable of accelerating the process of star formation.
One of the advantages of the ``magnetic'' scenario  is the possibility of explaining the observed composition
of the Radcliffe wave, which includes predominantly a gas--dust component and young stellar
objects. The gas and dust component frozen into the galactic magnetic field turns out to be most susceptible
to disturbance in the event of the development of Parker instability. The emerging magnetic field
arches pull gas and dust along with them, lifting them above the plane of the galactic disk. The movements of
the gas component along the magnetic field lines that
are, caused by gravitational forces lead to compactions
that facilitate star formation. The main components of
the disturbance in this scenario are the gas--dust environment
and young stellar objects (stars of early spectral
classes, young stars, and protostellar objects), the
age of which does not exceed the age of the Radcliffe
wave itself. The perturbation of the population of stars
formed before the onset of instability, on the contrary,
turns out to be insignificant. As a result, stars formed
before the development of instability will be located and move predominantly in the plane of the galactic disk. Young stellar objects formed after the development of instability in the gas filling the magnetic field arches will be located above the plane of the galactic disk and move under the action of gravity in the direction of the plane of symmetry of the disk. The age and velocity of movement of these objects make it possible to estimate the upper limit on the age of the Radcliffe wave itself and to limit the permissible time for the development of instability.

Estimation of the magnitude of the interstellar magnetic field strength $B$ in the Radcliffe wave zone can be obtained, taking into account that the kinetic energy of the material moving along the magnetic field lines, 
$E_{\rm k} \simeq (1/2) M_{\rm rw} v_{\rm rw}^2$, cannot exceed the value of magnetic energy in the Radcliffe wave, 
$E_{\rm m} \simeq (B^2/8 \pi) V_{\rm rw}$. This leads to the following inequality:
$B \geq B_0$, where
  \begin{equation}
  B_0 \simeq 2\, \mu G\,
   \left (\frac{M_{\rm rw}  }  {10^6\, M_\odot }  \right )^{1/2}
   \left (\frac {v_{\rm rw}}{10^6\, \hbox {\rm km/s}   }\right )
   \left (\frac {V_{\rm rw}}{10^{64}\, \hbox {\rm cm$^3$} }\right )^{-1/2}.
 \end{equation}
Here, $V_{\rm rw}$ is the volume of space, $M_{\rm rw}$ is the total mass of the substance, and $v_{\rm rw}$ is the velocity of movement of matter in the region of the Radcliffe wave. The estimate of the interstellar magnetic field strength that we
obtained appears to be rather realistic and corresponds to the values of this parameter previously obtained by other authors [70].

Finally, it should be noted that a possible cause (trigger) for the development of Parker instability
could be a local increase in the fraction of accelerated particles (cosmic rays) in a region of space with a characteristic
size of the order of several hundred pc. The most likely source of accelerated particles in this case
is a supernova explosion or series of supernova explosions (see, for example, [75] and reference therein).

\section{CONCLUSIONS}
Publications devoted to the study of the characteristics of the Radcliffe wave have been reviewed. The
advent of mass measurements of radial velocities of stars has recently led to a number of interesting results
obtained from the analysis of the spatial velocities of
young stars and, in particular, young open star clusters. An important place in the study has been given to issues related to the clarification of the direct or indirect influence of magnetic fields on the process of Radcliffe wave formation.

What is new is the establishment by various authors of radial motion in the galactic plane $XY$ front of dust
matter, molecular clouds and young stars. This Radcliffe
wave front is widely extended along the axis $y'$ by $\sim$2.5 kpc and is very narrow in the direction perpendicular
to this axis with a magnitude of $\sim$0.3 kpc. It appears that this motion is not associated with either
the galactic spiral density wave or other large-scale disturbances of the galactic disk.

Besides, the presence of vertical velocity disturbances $W$ in the Radcliffe wave has been indisputably
proven. The epicenter of disturbances in vertical coordinates and velocities is located in a relatively small
area of the Local System near the Sun, where the Gould Belt is located.

Estimates of the values of the main parameters and
manifestations of the Parker instability of the galactic
magnetic field support the hypothesis relating the formation
of the Radcliffe wave with instabilities of the
interstellar magnetic field in the galactic disk. A possible
trigger for such instability could be an increase in
the concentration of accelerated particles (cosmic
rays) in a local region of the galactic disk with a magnetic
field strength of several $\mu$G.

\subsection*{ACKNOWLEDGMENTS}
The authors are grateful to the reviewer for useful comments that contributed to improving the article.

\subsection*{FUNDING}
This work was supported by ongoing institutional funding.
No additional grants to carry out or direct this particular research were obtained.

\subsection*{CONFLICT OF INTEREST}
The authors of this work declare that they have no conflicts of interest.

\subsection*{REFERENCES}
\quad~ 1. H. Mineur, Mon. Not. R. Astron. Soc. 90, 789 (1930).

2. K. F. Ogorodnikov, Dynamics of Stellar Systems (Fizmatgiz, Moscow, 1965; Pergamon, Oxford, 2016).

3. Yu. N. Efremov, Star Formation Centers in Galaxies (Nauka, Moscow, 1989) [in Russian].

4. P. O. Lindblad, Bull. Astron. Inst. Netherland 19, 34 (1967).

5. P. O. Lindblad, K. Grape, A. Sandqvist, and J. Schober, Astron. Astrophys. 24, 309 (1973).

6. P. T. de Zeeuw, R. Hoogerwerf, J. H. J. de Bruijne,
A. G. A. Brown, and A. Blaauw, Astron. J. 117, 354 (1999).

7. J. Torra, D. Fern\'andez, and F. Figueras, Astron. Astrophys. 359, 82 (2000).

8. B. A. Gould, Proc. Am. Assoc. Adv. Sci., No. 1, 115 (1874).

9. B. A. Gould, Uranometria Argentina (P. E. Coni, Buenos Aires, 1879).

10. C. A. Olano, Astron. Astrophys. 121, 295 (2001).

11. V. V. Bobylev, Astrofizika 57, 625 (2014).

12. C. Zucker, J. S. Speagle, E. F. Schlafly, G. M. Green, D. P. Finkbeiner, A. Goodman, and J. Alves, Astron. Astrophys. 633, A51 (2020).

13. J. Alves, C. Zucker, A. A. Goodman, J. S. Speagle, et al., Nature (London, U.K.) 578 (7794), 237 (2020).

14. A. Blaauw, Koninkl. Ned. Akad. Wetenschap. 74 (4), 1 (1965).

15. R. Fleck, Nature (London, U.K.) 583 (7816), E24 (2020).

16. L. Thulasidharan, E. D'Onghia, E. Poggio, R. Drimmel,
J. S. Gallagher III, C. Swiggum, R. A. Benjamin, and J. Alves, Astron. Astrophys. 660, L12 (2022).

17. A. Marchal and P. G. Martin, Astrophys. J. 942 (2), 70 (2023).

18. R. Konietzka, A. A. Goodman, C. Zucker, A. Burkert, et al., Nature (London, U.K.) 628 (8006), 62 (2024).

19. W. G. L. P\"oppel, in From Darkness to Light: Origin and
Evolution of Young Stellar Clusters, Ed. by T. Montmerle and P. Andre, ASP Conf. Ser. 243, 667 (2001).

20. R. Lallement, J. L. Vergely, C. Babusiaux, and N. L. J. Cox, Astron. Astrophys. 661, A147 (2022).

21. G. Edenhofer, C. Zucker, P. Frank, A. K. Saydjari, J. S. Speagle, et al., Astron. Astrophys. 685, A82 (2024).

22. C. Zucker, J. Alves, A. Goodman, S. Meingast, and P. Galli, in Protostars and Planets VII, Proceedings of a
Conference, Kyoto, Japan, April 10--15, 2023, Ed. by S.-I. Inutsuka, Y. Aikawa, T. Muto, K. Tomida, and M. Tamura, ASP Conf. Ser. 534, 43 (2023).

23. V. V. Bobylev, A. T. Bajkova, and Yu. N. Mishurov, Astron. Lett. 48, 434 (2022).

24. G.-X. Li and B.-Q. Chen, Mon. Not. R. Astron. Soc. Lett. 517, L102 (2022).

25. J. Donada and F. Figueras, arXiv: 2111.04685 [astroph. GA] (2021).

26. V. V. Bobylev, N. R. Ikhsanov, and A. T. Bajkova, Astrophys. Byull. 80 (2), 181 (2025).

27. V. V. Bobylev, A. T. Baikova, and Yu. N. Mishurov, Astrofizika 65, 603 (2022).

28. M. Zhang, Astrophys. J. Suppl. 265, 59 (2023).

29. Z.-K. Zhu, M. Fang, Z.-J. Lu, J. Wang, et al., Astrophys. J. 971, 167 (2024).

30. G. V. Panopoulou, C. Zucker, D. Clemens, V. Pelgrims, et al., Astron. Astrophys. 694, A97 (2025).

31. J. Alonso-Santiago, A. Frasca, A. Bragaglia, G. Catanzaro, et al., Astron. Astrophys. 691, A317 (2024).

32. G.-X. Li, J.-X. Zhou, and B. Chen, Res. Not. Am. Astron. Soc. 8 (12), 316 (2024).

33. J. D. Soler, S. Molinari, S. C. O. Glover, R. J. Smith, et al., Astron. Astrophys. 695, A222 (2025).

34. C. Zucker, J. S. Speagle, E. F. Schlafly, G. M. Green,
D. P. Finkbeiner, A. A. Goodman, and J. Alves, Astrophys. J. 879, 125 (2019).

35. A. G. A. Brown, A. Vallenari, T. Prusti, J. H. J. de Bruijne, et al., Astron. Astrophys. 616, A1 (2018).

36. A. Vallenari, A. G. A. Brown, T. Prusti, J. H. J. de Bruijne, et al., Astron. Astrophys. 674, A1 (2023).

37. A. G. A. Brown, A. Vallenari, T. Prusti, J. H. J. de Bruijne, et al., Astron. Astrophys. 649, A1 (2021).

38. M. F. Skrutskie, R. M. Cutri, R. Stiening, M. D. Weinberg, et al., Astron. J. 131, 1163 (2006).

39. E. Winston, J. L. Hora, and V. Tolls, Astron. J. 160, 68 (2020).

40. M. A. Kuhn, R. S. de Souza, A. Krone-Martins, A. Castro-Ginard, et al., Astrophys. J. Suppl. 254, 33
(2021).

41. G. Marton, P. \'Abrah\'am, L. Rimoldini, M. Audard, et al., Astron. Astrophys. 674, A21 (2023).

42. G. Marton, P. \'Abrah\'am, E. Szegedi-Elek, J. Varga, et al., Mon. Not. R. Astron. Soc. 487, 2522 (2019).

43. V. V. Bobylev and A. T. Bajkova, Res. Astron. Astrophys.
24, 035010 (2024).

44. E. L. Hunt and S. Reffert, Astron. Astrophys. 673, A114
(2023).

45. T. Cantat-Gaudin, F. Anders, A. Castro-Ginard,
C. Jordi, et al., Astron. Astrophys. Suppl. Ser. 640, A1 (2020).

46. A. E. Piskunov, N. V. Kharchenko, S. R\"oser, E. Schilbach,
and R.-D. Scholz, Astron. Astrophys. 445, 545 (2006).

47. V. V. Bobylev, Astron. Lett. 46, 131 (2020).

48. M. J. Reid, K. M. Menten, A. Brunthaler, X. W. Zheng,
et al., Astrophys. J. 885, 131 (2019).

49. H. R. Dickel, H. J. Wendker, and J. H. Bieritz, in The
Spiral Structure of our Galaxy, Proc. 38th IAU Symposium,
Ed. by W. Becker and G. I. Kontopoulos, IAU
Symp. Proc. 38, 213 (1970).

50. Y. Xu, J. J. Li, M. J. Reid, K. M. Menten, et al., Astrophys.
J. 769, 15 (2013).

51. V. V. Bobylev and A. T. Bajkova, Astron. Lett. 40, 773
(2014).

52. A. V. Veselova and I. I. Nikiforov, Res. Astron. Astrophys.
20 (12), 209 (2023).

53. Y. Xu, C. J. Hao, D. J. Liu, Z. H. Lin, S. B. Bian,
L. G. Hou, J. J. Li, and Y. J. Li, Astrophys. J. 947, 54
(2023).

54. A. Siebert, B. Famaey, J. Binney, B. Burnett, et al.,
Mon. Not. R. Astron. Soc. 425, 2335 (2012).

55. C. Francis and E. Anderson, Mon. Not. R. Astron. Soc. 422, 1283 (2012).

56. J. P. Vall\'ee, Astron. J. 135, 1301 (2008).

57. V. V. Bobylev and A. T. Bajkova, Mon. Not. R. Astron.
Soc. 437, 1549 (2014).

58. M. J. Reid, K. M. Menten, X. W. Zheng, A. Brunthaler,
et al., Astrophys. J. 700, 137 (2009).

59. X. J. Shen, L. G. Hou, H. L. Liu, and X. Y. Gao,
Astron. Astrophys. 696, A67 (2025).

60. T. Antoja, P. Ramos, B. Garcia-Conde, M. Bernet,
C. F. P. Laporte, and D. Katz, Astron. Astrophys. 673,
A115 (2023).

61. C. Cao, Z.-Y. Li, R. Sch\"onrich, and T. Antoja, Astrophys. J. 975, 292 (2024).

62. E. Poggio, S. Khanna, R. Drimmel, E. Zari, et al., arXiv: 2407.18659 [astro-ph.GA] (2024).

63. R. Lallement, C. Babusiaux, J. L. Vergely, D. Katz,
F. Arenou, B. Valette, C. Hottier, and L. Capitanio,
Astron. Astrophys. 625, A135 (2019).

64. G.-X. Li, J.-X. Zhou, and B. Chen, Mon. Not. R.
Astron. Soc. Lett. 516, L35 (2022).

65. P. A. R. Ade, N. Aghanim, C. Armitage-Caplan,
M. Arnaud, et al., Astron. Astrophys. 571, A16 (2014).

66. L. Martinez-Medina, E. Poggio, and E. Moreno-Hilario,
arXiv: 2504.03843 [astro-ph.GA] (2025).

67. E. Maconi, J. Alves, C. Swiggum, S. Ratzenbock, et al.,
Astron. Astrophys. 694, A167 (2025).

68. X. Mai, B. Zhang, M. J. Reid, L. Moscadelli, et al, Astrophys. J. 949, 10 (2023).

69. V. V. Bobylev, Astron. Lett. 50, 796 (2024).

70. S. A. Kaplan and S. B. Pikelner, The Interstellar Medium
(Nauka, Moscow, 1979; Harvard Univ. Press, Harvard, 1970).

71. T. C. Mouschovias, M. W. Kunz, and D. A. Christie, Mon. Not. R. Astron. Soc. 397, 14 (2009).

72. E. Heintz, C. Bustard, and E. G. Zweibel, Astrophys. J. 891, 157 (2020).

73. D. Tharakkal, A. Shukurov, F. A. Gent, G. R. Sarson,
A. P. Snodin, and L. F. S. Rodrigues, Mon. Not. R. Astron. Soc. 525, 5597 (2023).

74. R. Habegger, E. G. Zweibel, and S. Wong, Astrophys.
J. 951, 99 (2023).

75. V. I. Romansky, A. M. Bykov, and S. M. Osipov, Adv. Space Res. 74, 4290 (2024).

\end{document}